\documentclass[english,nobib,notoc,justified,twocolumn,aps,prd]{revtex4-1}
\usepackage{amsmath}
\usepackage{mathptmx}
\usepackage{helvet}
\usepackage{newtxmath}
\usepackage[T1]{fontenc}
\usepackage[latin9]{inputenc}
\usepackage{graphicx}

\makeatletter

\usepackage{hyperref}

\makeatother

\usepackage{babel}
\begin{document}

\title{Effective containment explains sub-exponential growth in confirmed
cases of recent COVID-19 outbreak in Mainland China}

\author{Benjamin F. Maier}
\email{bfmaier@physik.hu-berlin.de}

\affiliation{Robert Koch Institute, Nordufer 20, D-13353 Berlin, Germany}

\author{Dirk Brockmann}

\affiliation{Robert Koch Institute, Nordufer 20, D-13353 Berlin, Germany}

\affiliation{Institute for Theoretical Biology, Humboldt-University of Berlin,
Philippstr. 13, D-10115 Berlin}

\date{\today}
\begin{abstract}
The recent outbreak of COVID-19 in Mainland China is characterized
by a distinctive algebraic, sub-exponential increase of confirmed
cases with time during the early phase of the epidemic, contrasting
an initial exponential growth expected for an unconstrained outbreak
with sufficiently large reproduction rate. Although case counts vary
significantly between affected provinces in Mainland China, the scaling
law $t^{\mu}$ is surprisingly universal, with a range of exponents
$\mu=2.1\pm0.3$. The universality of this behavior indicates that,
in spite of social, regional, demographical, geographical, and socio-economical
heterogeneities of affected Chinese provinces, this outbreak is dominated
by fundamental mechanisms that are not captured by standard epidemiological
models. We show that the observed scaling law is a direct consequence
of containment policies that effectively deplete the susceptible population.
To this end we introduce a parsimonious model that captures both,
quarantine of symptomatic infected individuals as well as population
wide isolation in response to mitigation policies or behavioral changes.
For a wide range of parameters, the model reproduces the observed
scaling law in confirmed cases and explains the observed exponents.
Quantitative fits to empirical data permit the identification of peak
times in the number of asymptomatic or oligo-symptomatic, unidentified
infected individuals, as well as estimates of local variations in
the basic reproduction number. The model implies that the observed
scaling law in confirmed cases is a direct signature of effective
contaiment strategies and/or systematic behavioral changes that affect
a substantial fraction of the susceptible population. These insights
may aid the implementation of containment strategies in potential
export induced COVID-19 secondary outbreaks elsewhere or similar future
outbreaks of other emergent infectious diseases. 
\end{abstract}

\maketitle

\section{Introduction}

The current outbreak of the new coronavirus in Mainland China (COVID-19,
previously named 2019-nCoV) is closely monitored by governments, researchers,
and the public alike \cite{cohen_scientists_2020,hsu_heres_2020,lewis_chinas_2020,who_novel_2020,chen_epidemiological_2020,zhao_preliminary_2020,cdc_2019_2020}.
The rapid increase of positively diagnosed cases in Mainland China
and subsequent exportation and confirmation of cases in more than
20 countries worldwide raised concern on an international scale. The
World Health Organization (WHO) therefore announced the COVID-19 outbreak
a Public Health Emergency of International Concern \cite{who_novel_2020}.

Confirmed cases in Mainland China increased from approx. 330 on Jan.~21st,
2020 to more than 17,000 on Feb.~2nd, 2020 in a matter of two weeks
\cite{center_for_systems_science_and_engineering_johns_hopkins_university_cssegisanddatacovid-19_2020}.
In Hubei Province, the epicenter of the COVID-2019 outbreak, confirmed
cases rose from 270 to 11,000 in this period, in all other Chinese
provinces the cumulated case count increased from 60 to 6,000 in the
same period.

\begin{figure*}
\includegraphics[width=1\textwidth]{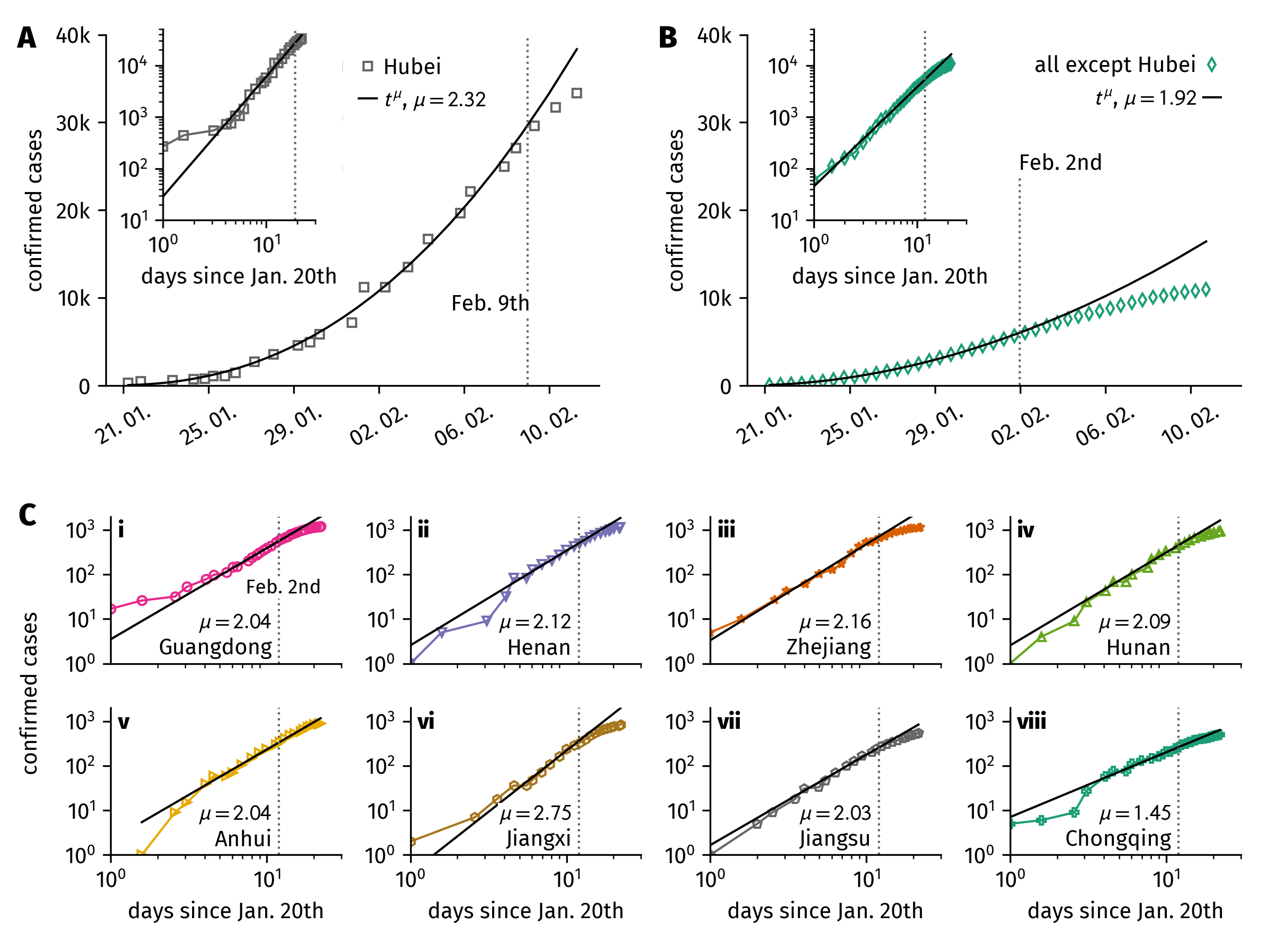}\caption{\textbf{\label{fig:data-china}A}: Confirmed case numbers in Hubei.
The increase in cases follows a scaling law $t^{\mu}$ with an exponent
$\mu=2.32$ after a short initial exponential growth phase. On Feb~9th
the case count starts deviating towards lower values. \textbf{B}:
Aggregated confirmed cases in all other affected provinces except
Hubei. $C(t)$ follows a scaling law with exponent $\mu=1.92$ until
Feb.~2nd when case counts deviate to lower values. The insets in
\textbf{A} and \textbf{B} depict $C(t)$ on a log-log scale. \textbf{C}:
Confirmed cases as a function of time for the 8 most affected provinces
in China. The curves follow a scaling law with exponents $\mu\approx2$
with the exception of Chongqing Province ($\mu=1.45$) and Jiangxi
Province ($\mu=2.75)$.}
\end{figure*}

An initial exponential growth of confirmed cases is generically expected
for an uncontrolled outbreak and in most cases mitigated with a time
delay by effective containment strategies and policies that reduce
transmission and effective reproduction of the virus, commonly yielding
a saturation in the cumulative case count and an exponential decay
in the number of new infections. Although in Hubei the number of cases
was observed to grow exponentially in early January \cite{li_early_2020},
the subsequent rise followed a sub-exponential, super-linear, algebraic
scaling law $t^{\mu}$ with an exponent $\mu=2.3$ (between Jan.~24th
and Feb.~9th), cf. Fig.~\ref{fig:data-china}A. For the majority
of the affected Chinese provinces of Mainland China, however, this
type of algebraic rise occured from the beginning, lacking an initial
exponential growth altogether. Surprisingly, the exponent $\mu$ does
not vary substantially with a typical value of $\mu=2.1\pm0.3$ for
the confirmed case curves in other substantially affected provinces
(confirmed case counts larger than 500 on Feb.~12th), despite geographical,
socio-economical differences, differences in containment strategies,
and heterogeneties that may have variable impacts on how the local
epidemic unfolds, cf. Fig.~\ref{fig:data-china}B~and~C.

The universality of the scaling relation $t^{\mu}$ with similar exponents
$\mu$ is evidence that this aspect of the dynamics is determined
by fundamental principles that are at work and robust with respect
to variation of other parameters that typically shape the temporal
evolution of epidemic processes. Three questions immediately arise,
(i) what may be the reason for this functional dependency, (ii) are
provinces other than Hubei mostly driven by export cases from Hubei
and therefore follow a similar functional form in case counts as suggested
by preliminary studies \cite{prasse_network-based_2020,sanche_novel_2020,chen_time_2020},
or, alternatively, (iii) is the scaling law a consequence of endogeneous
and basic epidemiological processes and a consequence of a balance
between transmission events and containment factors.

Here we propose a parsimonious epidemiological model that incorporates
quarantine measures and containment policies. In addition to standard
epidemiological parameters such as transmission and recovery rate,
the model introduces effective quarantine measures that act on \emph{both},
symptomatic infected individuals as well as susceptible individuals.
We show that depleting both ``ends'' of the transmission interactions
that involve an infectious and a susceptible individual naturally
yields a sub-exponential, algebraic increase in confirmed cases. This
behavior is qualitatively different from the behavior expected in
a scenario in which only infecteds are targets of containment strategies
to reduce transmission. In this case one expects either a stable exponential
growth at a lower rate or an exponential decay if mitigation is sufficiently
effective.

The model predicts the observed growth of case numbers exceptionally
well for almost all affected provinces for epidemiological parameters
estimated by preliminary studies \cite{who_novel_2020-1,cdc_2019_2020,sanche_novel_2020,zhao_preliminary_2020}.
Furthermore, the model permits an indirect assessment of the peak
time in the number of undetected infectious individuals (asymptomatic
and oligosymptomatic cases) that preceeds the depletion of confirmed
cases. This indicates that the number of unidentified infectious indviduals
peaked in early February.

\section{Situation assessment\label{sec:Situation-Assessment}}

\subsection{Case numbers in Hubei and other provinces\label{subsec:Case-numbers-in}}

We rely on case number data provided by the Systems Science and Engineering
group of Johns Hopkins University who currently give up to two updates
per day on the number of laboratory-confirmed cases globally \cite{center_for_systems_science_and_engineering_johns_hopkins_university_cssegisanddatacovid-19_2020}.
For the time period discussed in this paper, the group integrated
and curated case data manually collected at four sources: the World
Health Organization (WHO), the US Centers for Disease Control and
Prevention (CDC), the European Centre for Disease Prevention and Control
(ECDC), the Chinese physician's platform DXY.cn, and the National
Health Commission of the People's Republic of China (NHC). The published
data comprises total confirmed cases, total deaths, and total recovered
cases as a function of time for each affected location. In the following
we will focus on the number of total confirmed cases $C(t)$.

In Hubei, we find that the initital increase in confirmed cases is
followed by an algebraic scaling with time, i.e. $C(t)\propto t^{\mu}$,
with a scaling exponent $\mu\approx2.3$ that persisted until Feb.~9th,
see Fig.~\ref{fig:data-china}A. On Feb.~12th, the case definition
was changed by Chinese authorities which labeled a large number of
previously unconfirmed cases as confirmed, leading to a discontinuity
in the curves. We will therefore only consider data prior to Feb.~12th,
6am UTC.

Fig.~\ref{fig:data-china}B illustrates the cumulated case count
in all affected provinces except Hubei province. In the period Jan.~21st
until Feb.~2nd the curve follows an algebraic scaling law $t^{\mu}$
with $\mu\approx1.9$ lacking the initial exponential phase observed
in Hubei province. Starting on Feb.~2nd, the observed case count
starts deviating towards lower case counts. This poses the question
whether the observed behavior is an averaging affect introduced by
accumulating case counts across provinces. Interestingly, Fig.~\ref{fig:data-china}C
provides evidence that this is not so. The confirmed case counts in
the most affected provinces all exhibit a scaling law with exponents
close to $\mu=2$. Among the most affected provinces only Chongqing
Province exhibits a significantly lower exponent. Furthermore, all
provincial case count curves start deviating from the algebraic curve
around Feb.~2nd.

\subsection{Implemented policies\label{subsec:Implemented-Policies}}

The Chinese government put several mitigation policies in place to
contain the spread of the epidemic. In particular, positively diagnosed
cases were either quarantined or put under a form of self-quarantine
at home \cite{gan_over_2020}. Similarly, suspicious cases were confined
in monitored house arrest, e.g.~individuals who arrived from Hubei
before all traffic from its capital Wuhan was effectively restricted
\cite{fahrion_how_2020,rosales_wuhan_2020}. These measures aimed
at the removal of infectious individuals from the transmission process.

Additionally, measures were implemented that aimed at the protection
of the susceptible population by the partial shutdown of public life.
For instance, universities remained closed, many businesses closed
down, and people were asked to remain in their homes for as much time
as possible \cite{fahrion_how_2020,rosales_wuhan_2020,gan_over_2020}.
These actions that affect both, susceptibles and non-symptomatic infectious
individuals, not only protect susceptibles from acquiring the infection
but also remove a substantial fraction of the entire pool of susceptibles
from the transmission process and thus indirectly mitigate the proliferation
of the virus in the population in much the same way that herd immunity
is effective in the context of vaccine preventable diseases.

\section{Epidemiological modeling with quarantine and isolation\label{sec:modeling}}

On a very basic level, an outbreak as the one in Hubei is captured
by SIR dynamics \cite{keeling_modeling_2008}. The population is devided
into three compartments that differentiate the state of invididuals
with respect to the contagion process: (I)nfected, (S)usceptible to
infection, and (R)emoved (i.e. not taking part in the transmission
process). The corresponding variables $S$, $I$, and $R$ quantify
the respective compartments' fraction of the total population such
that $S+I+R=1$. The temporal evolution of the number of cases is
governed by two processes: The infection that describes the transmission
from an infectious to a susceptible individual with basic reproduction
number $R_{0}$ and the recovery of an infected after an infectious
period of average length $T_{I}$. The basic reproduction number $R_{0}$
captures the average number of secondary infections an infected will
cause before he or she recovers or is effectively removed from the
population.

Initially, a small fraction of infecteds yields an exponential growth
if the basic reproduction number is larger than unity. When the supply
of susceptibles is depleted, the epidemic reaches a maximum and the
infecteds decline. A simple reduction of contacts caused by isolation
policies could be associated with a reduction in the effective reproduction
number, which would, however, still yield an exponential growth in
the fraction of infecteds as long as $R_{0}>1$, inconsistent with
the observed scaling law $t^{\mu}$ discussed above. To test the hypothesis
that the observed growth behavior can be caused by isolation policies
that apply to both, infected and susceptible individuals, by effective
public shutdown policies, we extend the SIR model by two additional
mechanisms one of which can be interpreted as a process of removing
susceptibles from the transmission process. First, we assume that
general public containment policies or individual behavioral changes
in response to the epidemic effectively remove individuals from the
interaction dynamics or significantly reduce their participation in
the transmission dynamics. Secondly, we account for the removal of
symptomatic infected individuals by quarantine procedures. The dynamics
is governed by the system of ordinary differential equations:
\begin{align}
\partial_{t}S & =-\alpha SI-\kappa_{0}S\label{eq:dtS}\\
\partial_{t}I & =\alpha SI-\beta I-\kappa_{0}I-\kappa I\label{eq:dtI}\\
\partial_{t}X & =(\kappa+\kappa_{0})I,\label{eq:dtX}
\end{align}
a generalization of the standard SIR model, henceforth referred to
as the SIR-X model. The rate parameters $\alpha$ and $\beta$ quantify
the transmission rate and the recovery rate of the standard SIR model,
respectively. Additionally, the impact of public containment is captured
by the terms proportional to the containment rate $\kappa_{0}$ that
is effective in both $I$ and $S$ populations. Infected individuals
are removed at rate $\kappa$ corresponding to quarantine measures
that only affect symptomatic infecteds. The new compartment $X$ quantifies
symptomatic, quarantined infecteds. Here we assume that the fraction
$X(t)$ is proportional to the empirical confirmed and reported cases.
The case $\kappa_{0}=0$ corresponds to a scenario in which the general
population is unaffected by policies or does not commit behavioral
changes in response to an epidemic. The case $\kappa=0$ corresponds
to a scenario in which symptomatic infecteds are not isolated specifically.
Note that infecteds are always removed from their compartment $I$
at a higher rate than susceptibles as $\beta+\kappa+\kappa_{0}>\kappa_{0}$.

In the basic SIR model that captures unconstrained, free spread of
the disease, the basic reproduction number $R_{0}$ is related to
transmission and recovery rate by $R_{0}\equiv R_{0,\mathrm{free}}=\alpha/\beta$
because $\beta^{-1}=T_{I}$ is the average time an infected individual
remains infectious before recovery or removal. Here, the time period
that an infected individual remains infectious is $T_{I,\mathrm{eff}}=(\beta+\kappa_{0}+\text{\ensuremath{\kappa}})^{-1}$
such that the effective, or ``observed'' reproduction number $R_{0,\mathrm{eff}}=\alpha T_{I,\mathrm{eff}}$
is smaller than $R_{0,\mathrm{free}}$ since both $\kappa_{0}>0$
and $\kappa>0$.

We introduce \emph{public containment leverage}
\[
P=\frac{\kappa_{0}}{\kappa_{0}+\kappa}
\]
that reflects how strong isolation measures affect the general public
in comparison to quarantine measures imposed on symptomatic infecteds
alone. We further define
\[
Q=\frac{\kappa_{0}+\kappa}{\beta+\kappa_{0}+\kappa}
\]
as the \emph{quarantine probability}, i.e. the probability that an
infected is identified and quarantined either in specialized hospital
wards or at home.

The key mechanism at work in the model defined by Eqs.~(\ref{eq:dtS})-(\ref{eq:dtX})
is the exponentially fast depletion of susceptibles in addition to
isolation of infecteds. This effect is sufficient to account for the
observed scaling law in the number of confirmed cases for a plausible
range of rate parameters as discussed below.

\begin{figure*}[t]
\includegraphics[width=1\textwidth]{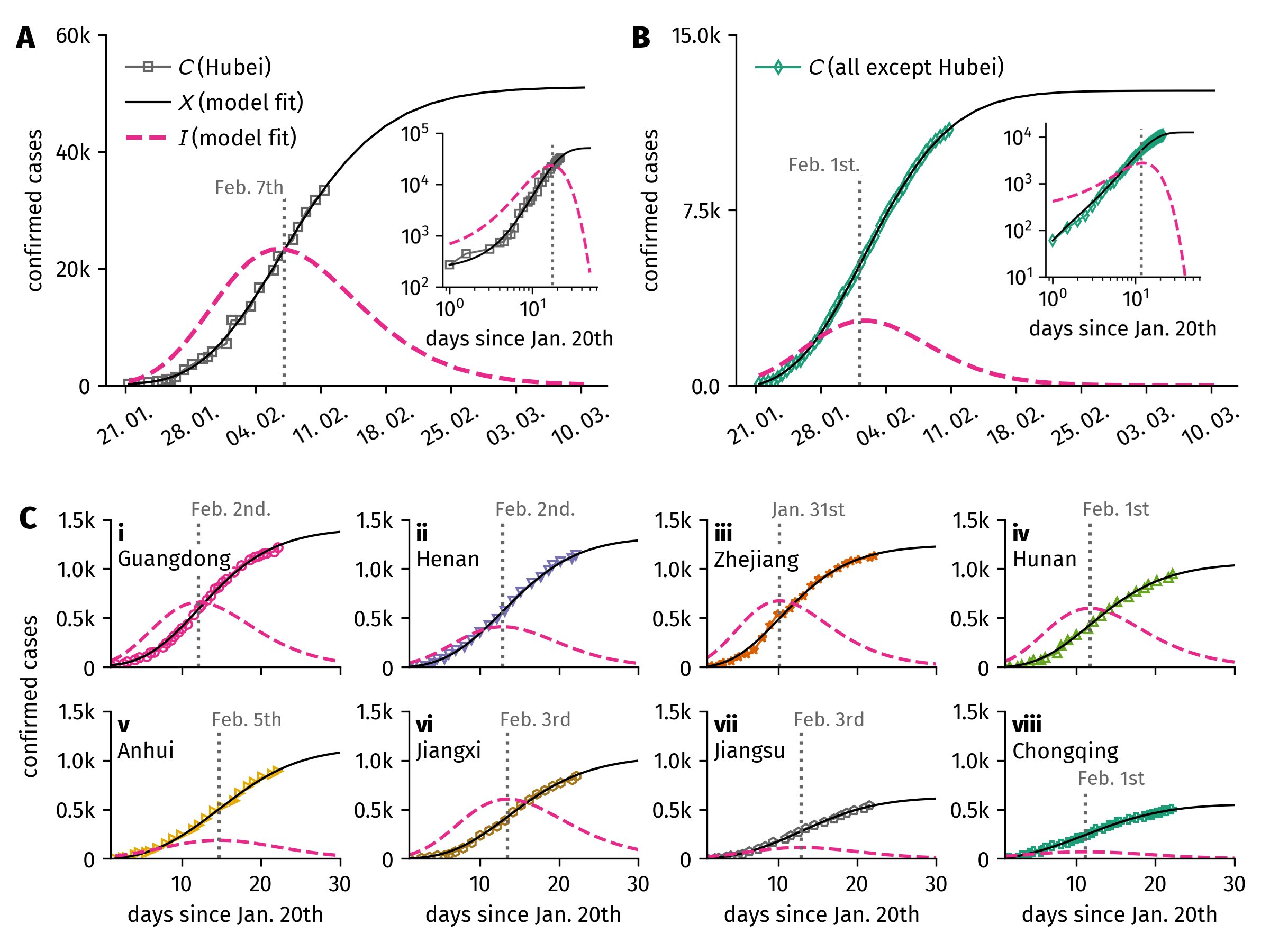}\caption{\label{fig:fit_mainland_china}Case numbers in Hubei compared to model
predictions. The quarantined compartment $X(t)$ and the unidentified
infectious compartment $I(t)$ are obtained from least-squares fits
of the model defined by Eqs.~(\ref{eq:dtS})-(\ref{eq:dtX}), cf.
App. \ref{app:Fitting-routine}. All fits were performed with fixed
values of $R_{0,\mathrm{free}}=6.2$ and $T_{I}=8\mathrm{d}$. Note
that the effective, observed basic reproduction number $R_{0,\mathrm{eff}}$
has lower values and varies for each of the affected provinces as
discussed in Sec.~\ref{sec:Results} and shown in Tab.~\ref{tab:Fit-parameters.}.
\textbf{A:} In Hubei, the model captures both, the initial rise of
confirmed cases as well as the subsequent algebraic growth. The confirmed
cases are predicted to saturate at $C=51,000$. The model also predicts
the time-course of the number of unidentified infectious individuals
$I(t)$ which peaks on Feb.~7th and declines exponentially afterwards.
While the order of $I(t)$ is associated with rather large fluctuations
depending on the fitting procedure, the predicted peak time is robust,
consistently around Feb.~7th. \textbf{B}: Model prediction for case
numbers aggregated over all affected provinces other than Hubei. The
case numbers' algebraic growth is well reflected and predicted to
saturate at $C=12,600$. In contrast to Hubei, the fraction of unidentified
infecteds peaks around Feb.~1st, approximately a week earlier. The
insets in \textbf{A} and \textbf{B} depict both data and fits on a
log-log scale. \textbf{C}: Fits for confirmed cases as a function
of time for the remaining 8 most affected provinces in China. All
curves are well captured by the model fits that predict similar values
for the peak time of unidentified infecteds.}
\end{figure*}

\section{Results\label{sec:Results}}

We assume that a small number of infected individuals travelled from
Hubei to each of the other affected provinces before traffic restrictions
were effective but at a time when containment measures were just being
implemented. Fig.~\ref{fig:fit_mainland_china} illustrates the degree
to which the case count for Hubei province and the aggregated case
count for all other provinces is captured by the SIR-X model as defined
by Eqs.~(\ref{eq:dtS})-(\ref{eq:dtX}).
\begin{table}
\begin{tabular}{lrrrrr} \hline  Province      &   $N/10^6$ &   $Q$ &   $P$ &   $I_0/X_0$ &   $R_{0,\mathrm{eff}}$ \\ \hline  Hubei         &       57.1 &  0.47 &  0.66 &        2.55 &                   3.28 \\  All exc. Hubei &     1277.8 &  0.69 &  0.21 &        6.98 &                   1.93 \\  Guangdong     &      104.3 &  0.51 &  0.75 &        3.66 &                   3.02 \\  Henan         &       94.3 &  0.61 &  0.38 &       41.00 &                   2.41 \\  Zhejiang      &       51.2 &  0.50 &  0.98 &       18.83 &                   3.10 \\  Hunan         &       67.0 &  0.47 &  1.00 &       58.34 &                   3.28 \\  Anhui         &       64.6 &  0.73 &  0.12 &       27.46 &                   1.70 \\  Jiangxi       &       44.0 &  0.44 &  1.00 &       19.41 &                   3.46 \\  Jiangsu       &       76.8 &  0.72 &  0.15 &       19.10 &                   1.75 \\  Chongqing     &       28.4 &  0.78 &  0.07 &        5.17 &                   1.36 \\ \hline \end{tabular}\caption{\label{tab:Fit-parameters.}Fit parameters as described in Sec.~\ref{sec:modeling},
fixed population size $N$, and resulting effective basic reproduction
number $R_{0,\mathrm{eff}}$ for Hubei and the remaining majorly affected
provinces discussed in the main text, decreasingly ordered by largest
case number. The infectious period $T_{I}$ and the basic reproduction
number $R_{0,\mathrm{free}}=\alpha T_{I}$ were fixed to values $T_{I}=8\,\mathrm{d}$
and $R_{0,\mathrm{free}}=6.2$, respectively.}
\end{table}

We find that for a wide range of model parameters, the case count
is well reproduced by the model. The model reproduces the scaling
law $t^{\mu}$ as observed in the data for a significant period of
time before saturating to a finite level. Remarkably, the model is
able to reproduce both growth behaviors observed in the data: The
model predicts the expected initial growth of case numbers in Hubei
Province followed by an algebraic growth episode for $\approx11$
days until the saturation sets in, a consequence of the decay of unidentified
infected individuals after a peak time around Feb.~7th (see Fig.~\ref{fig:fit_mainland_china}A).
Furthermore, the model also captures the immediate sub-exponential
growth observed in the remaining most affected provinces (Fig.~\ref{fig:fit_mainland_china}B-C).
Again, saturation is induced by a decay of unidentified infecteds
after peaks that occur several days before peak time in Hubei, ranging
from Jan.~31st to Feb.~5th. For all provinces, following their respective
peaks the number of unidentified infecteds $I(t)$ decays over a time
period that is longer than the reported estimation of maximum incubation
period of $14$ days \cite{cdc_2019_2020,who_novel_2020-1}. It is
important to note that due to the uncertainty in the population size,
the numerical value of unidentified infecteds is sensitive to parameter
variations---the general shape of $I(t)$, however, is robust for
a wide choice of parameters, as discussed in App.~\ref{app:Fitting-routine}.

Parameter choices for best fits were a fixed basic reproduction number
of $R_{0,\mathrm{free}}=6.2$ (note that this reproduction number
corresponds to an unconstrained epidemic) and a fixed mean infection
duration of $T_{I}=8\,\mathrm{d}$ consistent with previous reports
concerning the incubation period of COVID-19 \cite{who_novel_2020-1,cdc_2019_2020}.
The remaining fit parameters are shown in Tab.~\ref{tab:Fit-parameters.}.
For these values, the effective basic reproduction number is found
to range between $1.7\leq R_{0,\mathrm{eff}}\leq3.3$ for the discussed
provinces, consistent with estimates found in previous early assessment
studies \cite{cdc_2019_2020,kucharski_early_2020,read_novel_2020,zhao_preliminary_2020}.

A detailed analysis of the obtained values for quarantine probability
$Q$ and public containment leverage $P$ indicates that a wide range
of these parameters can account for similar shapes of the respective
case counts. Consequently, the model is structurally stable with respect
to these parameters and the numerical value is of less importance
than the quality of the mechanism they control (see App.~\ref{app:Fitting-routine}).

For the remaining 20 provinces, we find growth behaviors similar to
the ones discussed before. Some follow a functional form comparable
to the case counts in Hubei, others display a stronger agreement with
pure algebraic growth, as can be seen in Fig.~\ref{fig:all_provinces}.
Generally, a stronger agreement with algebraic growth can be associated
with stronger public containment leverages.

In order to stress the general importance of public containment measures
and to clarify to what extent quarantine is connected to case confirmation
we also analyzed a model variant where (i) the general public and
infecteds are removed from the transmission process in the same way
and (ii) case counting is decoupled from the quarantine process (see
App.~\ref{app:Case-confirmation-without-quarantine}). We find that
these model variants describe the real case count data reasonably
well for the majority of provinces, further evidence of the importance
of containment policies that target the susceptible population. In
App.~\ref{app:Case-number-development-in} we present further analytical
evidence for this conclusion.

Note that the described saturation behavior of confirmed cases requires
that eventually all susceptibles will effectively be removed from
the transmission process. In reality, not every susceptible person
can be isolated for such an extended period of time as the model suggests.
One might expect instead that the number of infecteds will decay more
slowly and saturate to a small, yet non-zero level. At this point
it would be crucial to identify unquarantined infecteds more efficiently
in order to completely shut down the transmission process. Due to
potential difficulties in upholding the containment policies for such
an extended period of time, we expect that our predictions will underestimate
the final total amount of confirmed cases.

\section{Discussion and Conclusion\label{sec:Conclusion}}

In summary, we find that one of the key features of the dynamics of
the COVID-19 epidemic in Hubei Province but also in all other provinces
is the robust sub-exponential rise in the number of confirmed cases
according to the scaling law $t^{\mu}$ during the first episode of
the epidemic. This general, almost universal behavior is evidence
that fundamental principles are at work associated with this particular
epidemic that are dominated by the interplay of the contagion process
with endogeneous behavioral changes in the susceptible population
and external mitigation policies. We find that the generic scaling
of case counts with time is independent of many other factors that
often shape the time-course of an epidemic.

The model defined by Eqs.~(\ref{eq:dtS})-(\ref{eq:dtX}) and discussed
here indicates that this type of behavior can generally be expected
if the supply of susceptible individuals is systematically decreased
by means of implemented containment strategies or behavioral changes
in response to information about the ongoing epidemic. Unlike contagion
processes that develop without external interference at all or processes
that merely lead to parametric changes in the dynamics, our analysis
suggests that non-exponential growth is expected when the supply of
susceptibles is depleted on a timescale comparable to the infectious
period assiociated with the infection. The effective depletion of
the susceptible populations on a timescale $\kappa_{0}^{-1}$ in the
model does not imply that the pool of susceptibles must be depleted
in a physical sense. Rather, this depletion could be achieved indirectly.

Containment that targets the susceptible population is a requirement
for the observed algebraic scaling of case counts with time, unlike
the reduction of transmission rates, or the reduction of the duration
infected individuals remain infectious. Both of the latter measures
address infected individuals only, which leads to a reduction of the
exponential growth rate of new infections, or an exponential decay
if the corresponding measures are sufficiently effective, but do not
alter the functional growth behavior.

The fact that scaling laws were observed in Hubei after an initial
exponential phase and that an exponential phase was not observed in
the other substantially affected provinces is consistent with the
model predictions if one assumes that the response in the population
and the effect of policies lagged the onset of the epidemic in Hubei
province and was immediate in all other provinces that were in an
alert status already when cases were increasing substantially in Hubei
province.

The model reproduces the empirical case counts in all provinces well
for plausible parameter values. The quality of the reproduction of
the case counts in all 29 affected provinces can be used to estimate
the peak time of the number of asymptomatic or oligo-symptomatic infected
individuals in the population, which is the key quantity for estimating
the time when this outbreak will wane. The current analysis indicates
that this peak time was reached around Feb.~7th for Hubei and within
the first days of February in the remaining affected provinces.

The model further suggests that the public response to the epidemic
and the containment measures put in place by the Chinese administration
were effective despite the increase in confirmed cases. That this
behavior was universally observed in all provinces also indicates
that mitigation strategies were universally effective. Based on our
analysis, we argue that the implemented containment strategies should
stay in effect for a longer time than the incubation period after
the saturation in confirmed cases sets in for this particular outbreak.

Our analysis further implies and provides evidence that mitigation
strategies that target the susceptible population and induce behavioral
changes at this ``end'' of the transmission process are most effective
to contain an epidemic---especially in situations when asymptomatic
or mildly symptomatic infectious periods are long or their duration
unknown. This may be of importance for developing containment strategies
in future scenarios or if the current COVID-19 epidemic were to trigger
large scale outbreaks in other regions of the world by exports and
subsequent proliferation.
\begin{acknowledgments}
We want to express our gratitude to L\@.~H.~Wieler and L.~Schaade
for helpful comments regarding the manuscript. B.~F.~M.~is financially
supported as an \emph{Add-on Fellow for Interdisciplinary Life Science
}by the Joachim Herz Stiftung.
\end{acknowledgments}

\appendix

\section{Fitting routine and analysis\label{app:Fitting-routine}}

Fits were performed using the Levenberg-Marquardt method of least
squares. We fixed the epidemiological parameters to duration of infection
$T_{I}=8\,\mathrm{d}$ and basic reproduction number $R_{0,\mathrm{free}}=\alpha T_{I}=6.2$.
The population size $N$ of each of the affected Chinese provinces
was obtained from the Geonames project \cite{geonames_geonames_2019}
and is listed in Tab.~\ref{tab:Fit-parameters.}. For each confirmed
cases data set, we set the initial conditions $X(t_{0})=C(t_{0})/N$,
$I(t_{0})=(I_{0}/X_{0})X(t_{0})$, and $S(t_{0})=1-I(t_{0})-X(t_{0})$
where $t_{i},C(t_{i})$ is the $i$-th pair of a province's time and
aggregated confirmed case number. Since the number of unidentified
infectious is unknown per definition, the prefactor $I_{0}/X_{0}\geq1$
was chosen as a fit parameter. The remaining fit parameters were quarantine
rate $\kappa>0$ and containment rate $\kappa_{0}>0$. For the fit
procedure, Eqs.~(\ref{eq:dtS})-(\ref{eq:dtX}) were integrated using
the Dormand--Prince method which implements a fourth-order Runge--Kutta
method with step-size control, yielding $I(t)$ and $X(t)$ for every
parameter configuration and every data set. The residuals were computed
as $\mathcal{R}(t_{i})=NX(t_{i})-C(t_{i}).$

We find that the model accurately reflects both the sub-exponential
growth as well as the saturating behavior observed in the data discussed
in the main text, with the obtained fit parameters displayed in Tab.~\ref{tab:Fit-parameters.}.
For the 9 majorly affected provinces and the aggregated data over
all provinces except Hubei, the quarantine probability is found to
have similar values of $Q=0.6\pm0.1$. The public containment leverage
is fluctuating more strongly with values ranging between $P=0.07$
and $P=1$, where generally, lower values appear concurrently with
a higher quarantine probability, which suggests that stronger quarantine
implementation requires less public isolation. Yet, we advise not
to overinterpret the values of these fit parameters, as many different
parameter values generate similar developments of confirmed cases.

As we fixed the population size to be equal to the total population
of the respective provinces, one might also ask how system size changes
the results---for instance, if the outbreak is contained in a small
region of a province, the effective population available to the transmission
process will be substantially smaller. Therefore, we repeated the
fit procedure with $N$ as a free fit parameter to find that the form
of $X(t)$ is not altered substantially for different values of $N$,
while $I(t)$ can vary more strongly as reflected by a larger variation
in $I_{0}$. This result suggests that the estimation of the number
of infecteds is associated with a larger uncertainty. The general
shape of $I(t)$, however, remains stable such that an inference of
the peak time of unidentified infecteds is reasonable. Furthermore,
the model favours larger values of the containment rate $\kappa_{0}$
for larger population sizes in order to reproduce emprical data. This
is a reasonable because the number of available transmission pathways
grows quadratically with the population size, making it easier for
the disease to spread faster. In order to still observe sub-exponential
growth, a large part of the susceptible population has to be removed
quickly. Consistently, we find that a decay of susceptibles is necessary
to reliably obtain the observed growth behavior (i.e.~$P>0$, $\kappa_{0}>0$).

Concerning variation in the fixed model parameters, larger values
of up to $R_{0,\mathrm{free}}=12$ yield results similar to the ones
described above when adjusting the infectious period $T_{I}$ to larger
values, as well. For lower values of $R_{0,\mathrm{free}}<6$, the
model fails to reproduce the scaling laws observed in the data. Similarly,
the fit results are reasonably robust against variations of the duration
of infection in a range of $T_{I}\in[6\,\mathrm{d},20\,\mathrm{d}]$
with concurrent adjustment of $R_{0,\mathrm{free}}$.

\begin{figure*}[t]
\includegraphics[width=1\textwidth]{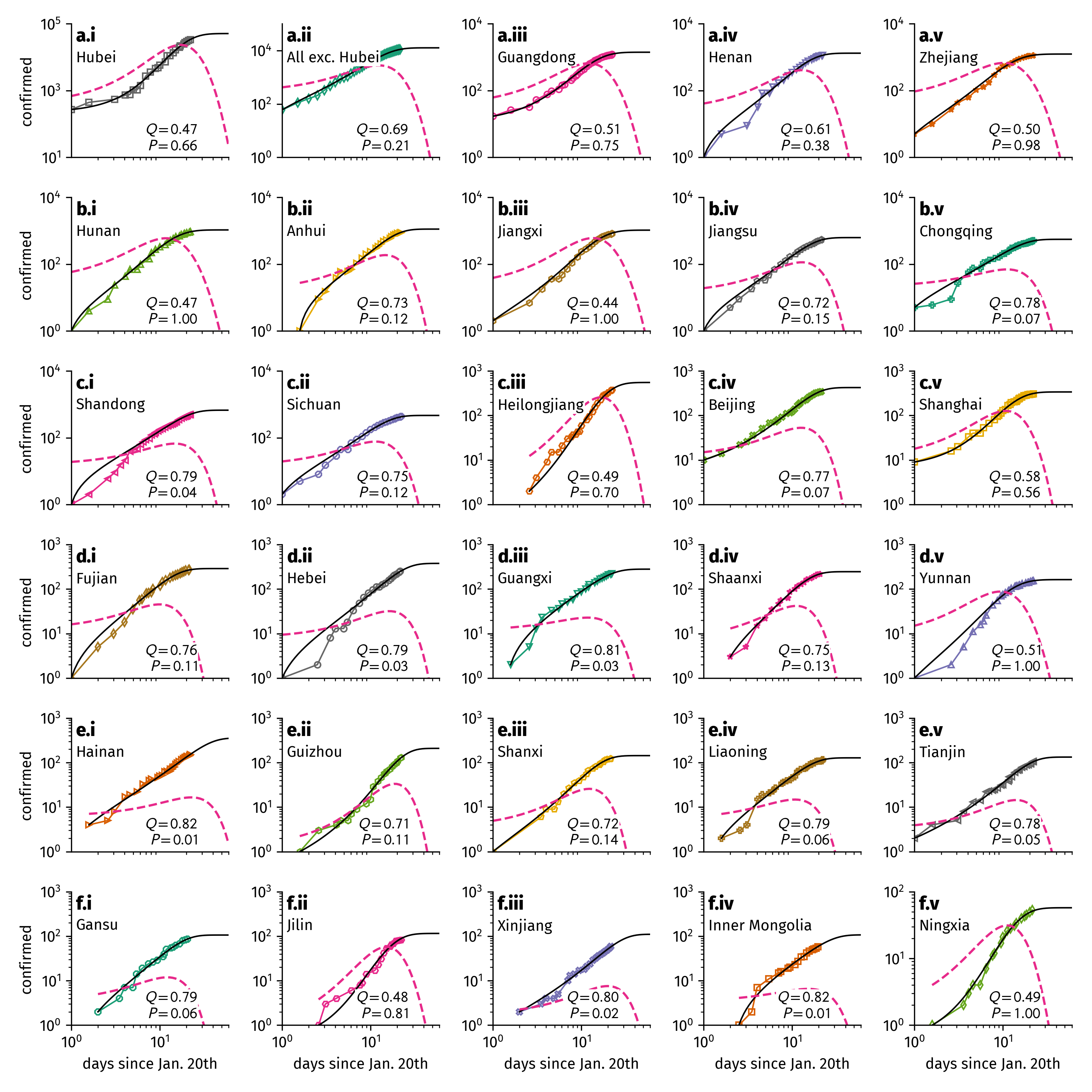}\caption{\label{fig:all_provinces}Case numbers of all affected provinces reproduced
by the SIR-X model defined by Eqs.~(\ref{eq:dtS})-(\ref{eq:dtX})
and the fraction of unidentified infecteds $I(t)$ as obtained from
least-squares fits. All fits were performed with fixed values of $R_{0,\mathrm{free}}=6.2$
and $T_{I}=8\mathrm{d}$. For more details concerning the fitting
procedure, cf.~App.~\ref{app:Fitting-routine}.}
\end{figure*}

Similar effects are found for the remaining 20 affected provinces
for which model fits are displayed in Fig.~\ref{fig:all_provinces}
with parameters given in Tab.~\ref{tab:Fit-parameters.-2}. In general,
provinces with larger values of public containment leverage $P$ exhibit
a stronger agreement with the hypothesis of algebraic growth.

All data and the analysis material is available online \cite{maier_benmaiercovid19casenumbermodel_2020}.

\begin{table}
\begin{tabular}{lrrrrr} \hline  Province       &   $N/10^6$ &   $Q$ &   $P$ &   $I_0/X_0$ &   $R_{0,\mathrm{eff}}$ \\ \hline  Shandong       &       94.2 &  0.79 &  0.04 &       18.86 &                   1.33 \\  Sichuan        &       87.2 &  0.75 &  0.12 &        9.74 &                   1.58 \\  Heilongjiang   &       38.2 &  0.49 &  0.70 &        6.13 &                   3.19 \\  Beijing        &       14.9 &  0.77 &  0.07 &        1.50 &                   1.42 \\  Shanghai       &       22.3 &  0.58 &  0.56 &        1.99 &                   2.58 \\  Fujian         &       36.9 &  0.76 &  0.11 &       16.10 &                   1.47 \\  Hebei          &       69.9 &  0.79 &  0.03 &        9.41 &                   1.27 \\  Guangxi        &       48.2 &  0.81 &  0.03 &        6.83 &                   1.15 \\  Shaanxi        &       37.6 &  0.75 &  0.13 &        4.33 &                   1.55 \\  Yunnan         &       45.4 &  0.51 &  1.00 &       14.99 &                   3.01 \\  Hainan         &        9.3 &  0.82 &  0.01 &        1.77 &                   1.12 \\  Guizhou        &       37.9 &  0.71 &  0.11 &        2.23 &                   1.81 \\  Shanxi         &       34.1 &  0.72 &  0.14 &        4.93 &                   1.71 \\  Liaoning       &       43.1 &  0.79 &  0.06 &        3.51 &                   1.28 \\  Tianjin        &       14.0 &  0.78 &  0.05 &        1.96 &                   1.37 \\  Gansu          &       26.3 &  0.79 &  0.06 &        2.50 &                   1.30 \\  Jilin          &       27.3 &  0.48 &  0.81 &        3.79 &                   3.22 \\  Xinjiang       &       21.3 &  0.80 &  0.02 &        1.13 &                   1.22 \\  Inner Mongolia &       24.3 &  0.82 &  0.01 &        4.11 &                   1.11 \\  Ningxia        &        6.2 &  0.49 &  1.00 &        3.99 &                   3.16 \\ \hline \end{tabular}\caption{\label{tab:Fit-parameters.-2}Fit parameters as described in Sec.~\ref{sec:modeling},
fixed population size $N$, and resulting effective basic reproduction
number $R_{0,\mathrm{eff}}$ for the remaining affected provinces,
decreasingly ordered by largest case number. The infectious period
$T_{I}$ and the basic reproduction number $R_{0,\mathrm{free}}=\alpha T_{I}$
were fixed to values $T_{I}=8\,\mathrm{d}$ and $R_{0,\mathrm{free}}=6.2$,
respectively.}
\end{table}

\section{Case-number development in early phase of the epidemic\label{app:Case-number-development-in}}

In a typical outbreak scenario only few people are infected initially,
such that $S(0)=1-\varepsilon$ and $I(0)=\epsilon$ with $\varepsilon\ll1$,
which implies that the depletion of susceptibles available to the
transmission process will be dominated by shutdown policies. This
assumption effectively linearizes Eq.~(\ref{eq:dtS}), yielding the
solution
\[
S(t)=\exp(-\kappa_{0}t).
\]
Further integration of the linearized Eq.~(\ref{eq:dtI}) yields
\[
I(t)=I_{0}\exp\left[-(\beta+\kappa+\kappa_{0})t\right]\times\exp\left[-\frac{\alpha}{\kappa_{0}}\left(\mathrm{e}^{-\kappa_{0}t}-1\right)\right].
\]
For small values of $t$ we can expand the exponent to obtain the
approximate growth function
\begin{align*}
I(t)/\tilde{I}_{0} & \approx\exp\left\{ \left(\alpha-\beta-\kappa-\kappa_{0}\right)t\right\} \times\exp\left\{ -\frac{1}{2}\alpha\kappa_{0}t^{2}\right\} 
\end{align*}
Here, the first factor implies that quarantining infecteds merely
decreases the rate with which their number will grow exponentially,
while the second term suppresses the whole transmission process well
enough to alter the growth behavior. This implies that public shutdown
policies facilitate epidemic containment in a more effective way than
quarantine measures.

\section{Case confirmation without quarantine\label{app:Case-confirmation-without-quarantine}}

In an alternative approach to the one discussed in the main text one
may assume that isolation affects all citizens equally, i.e.~$\kappa=0$,
but that the confirmation and counting of a new case is decoupled
from whether the respective person is quarantined or not---rather,
an infected is discovered and counted at rate $\tilde{\kappa}$. This
implies an ODE system of
\begin{align}
\partial_{t}S & =-\alpha SI-\kappa_{0}S\label{eq:dtS-1}\\
\partial_{t}I & =\alpha SI-\beta I-\kappa_{0}I\label{eq:dtI-1}\\
\partial_{t}X & =\tilde{\kappa}I,\label{eq:dtX-1}
\end{align}
where as before, $X$ will be equal to the number of confirmed cases
$C$.

A numerical investigation of this alternative model shows that confirmed
case data can be reasonably fit for all provinces, as well, which
confirms the argumentation of App.~\ref{app:Case-number-development-in}
that removal of susceptibles from the transmission process is more
effective than quarantine in mitigating epidemic spread.

\bibliography{ms.bbl}

\end{document}